# The structure and local chemical properties of boron-terminated tetravacancies in hexagonal boron-nitride

Ovidiu Cretu[*], Yung-Chang Lin, Masanori Koshino, Luiz H. G. Tizei, Zheng Liu, Kazutomo Suenaga

National Institute of Advanced Industrial Science and Technology (AIST), Nanotube Research Center, Central 5, 1-1-1 Higashi, Tsukuba, Ibaraki 305-8565, Japan

81.05.Ea, 61.72.-y, 68.37.Ma, 79.20.Uv

**Imaging and spectroscopy performed in a low-voltage scanning transmission electron microscope (LV-STEM) are used to characterize the structure and chemical properties of boron-terminated tetravacancies in hexagonal boron nitride (h-BN). We confirm earlier theoretical predictions about the structure of these defects and identify new features in the electron energy-loss spectra (EELS) of B atoms using high resolution chemical maps, highlighting differences between these areas and pristine sample regions. We correlate our experimental data with calculations which help explain our observations.**

Hexagonal BN has gained a lot of attention following a large amount of work on graphene [1], which has led to the discovery of other 2D graphene-analogues [2]. Its properties have been recently reviewed in [3]. Although similar to graphene in terms of crystal structure, the electronic behavior of h-BN is completely different, as the material is a wide (~6 eV) gap semiconductor. When compared to large number of studies on defects in graphene, there are relatively few similar reports for h-BN. Although some electron microscopy studies have focused on the structure of point defects [4-8] and edges [5, 9, 10], there is a lack of experimental data concerning the local chemical properties of defects.

Aberration-corrected LV-STEM has already demonstrated the ability to resolve both the structure and chemical properties of materials at the atomic scale, even in the case of delicate single-layered samples. The reduced specimen knock-on damage due to the lower acceleration voltage has allowed local chemical characterization even in areas which are more easily damaged, such as the edges of graphene [11, 12] and h-BN [10]. Additionally, single defect EELS spectroscopy has already been reported, both in the case of graphene [13, 14] and h-BN [7].

---

[*]ovidiu.cretu@aist.go.jp

In this work, we use LV-STEM in order to provide the first report of the structure and local chemical properties of boron-terminated tetravacancies in h-BN. The first part of the paper describes the atomic structure of the defects. The second part details anomalies observed in the EELS signature of B atoms. Finally, the third part of the results shows calculations which explain the structure and the newly-observed EELS spectral features.

The samples studied in this work were prepared using monolayer h-BN grown on Cu foil by chemical vapor deposition. The h-BN surface was first protected by spin coating with 1.5 μm thick 1wt% polycarbonate. The Cu foil was etched using diluted HCl, and then the single-layer h-BN was transferred onto the TEM grid. The specimens were cleaned overnight using chloroform before inserting in the microscope [15, 16]. Throughout the course of the experiments, the samples were repeatedly heated while inside the microscope, under vacuum, at temperatures up to 600 °C. High-resolution imaging and chemical analysis were performed using a Jeol JEM-2100F microscope, fitted with a cold field-emission gun, dodecapole-based aberration correctors (Jeol) and electron energy spectrometer (Gatan Quantum), operated at 60 and 30 kV [17]. Additional imaging was performed with a Jeol ARM-200F microscope, fitted with a Shottky-type field-emission gun and hexapole-based aberration corrector (Ceos), operated at 80 kV. The acquired EELS data was processed by PCA filtering, as implemented in the MSA plug-in [18] (HREM Research) for Digital Micrograph (Gatan). Additional processing was performed using ImageJ [19] and the Cornell Spectrum Imager [20].

In order to simulate EELS spectra for various configurations of h-BN, we adopted a band-structure calculation method involving pseudo-potentials. More details regarding the simulations can be found in the Supplementary Material [21]. We employed the density functional theory (DFT) based CASTEP module in Materials Studio v7.0 (Accelrys Co.), using the PBE [22] functional of the generalized gradient approximation (GGA). A core-hole was introduced for core spectroscopy analysis. The obtained spectra are smeared with 0.3 eV of energy broadening (FWHM) with a consideration of lifetime effect.

A high-resolution STEM image of a tetravacancy in h-BN is shown in Fig. 1. The signal-to-noise (S/N) ratio was improved by overlapping several consecutive aligned frames. The incoherent annular dark-field (ADF) image can be directly interpreted in order to determine the structure of the defect, as the intensity is proportional to the atomic number of the imaged atom, allowing us to distinguish between B and N. The corresponding atomic model is drawn on the right-hand side of the image and represents a $V_{B1N3}$ defect, obtained by removing one B and three N atoms from the perfect lattice. Importantly, while this structure is stable under our imaging conditions for time intervals sufficient in order to acquire image and spectral data (typically <60 s), it always evolves into larger defects. In our experience, the stability of the larger defect structures decreases with defect size, leading to the destruction of the area of the sample under the beam.

The average distance between two under-coordinated B atoms (marked $d_{BB}$ in the figure) is 2.02 ± 0.1 Å which is smaller than the value estimated in the bulk region of 2.50 Å (after re-calibrating the image using an unperturbed B-N bond length of 1.44 Å). This is in good agreement with [23], which studied this defect theoretically. The authors discussed both "normal" and "electron-rich" conditions, yet our experimental distance is closer to their 1.95 Å "normal" case, which is to be expected, as thin samples acquire (somewhat counterintuitively) a positive charge while being observed by TEM [24].

The stability of our $V_{B1N3}$ defects is in an apparent contradiction with previous literature reports, which have predominantly observed the sister $V_{B3N1}$ defect by examining h-BN using 120 kV [4] and 80 kV [5, 6] electrons in parallel-beam TEM mode at room temperature. In order to investigate this discrepancy, we looked at the types of defects created using various imaging conditions. The results are summarized in Table I.

TABLE I. h-BN defect types as a function of imaging conditions. The data in the shaded cells is taken from the literature.

|  | **Voltage** | 80 kV | | 60 kV |
|---|---|---|---|---|
|  | **Temperature** | 20 °C | 500 °C | 500 °C |
| **Mode** | TEM | $V_{B3N1}$ [4-6] |  | $V_{B3N1}$ $V_{B1N3}$ |
|  | STEM | $V_{B3N1}$ | $V_{B1N3}$ | $V_{B1N3}$ |

At **80 kV**, in STEM mode, we have found that the type of defects which are created depends on the sample temperature. When imaging at room temperature, B vacancies and $V_{B3N1}$ tetravacancies form predominantly. This is in agreement with the previous observations. However, at 500 °C, N vacancies and $V_{B1N3}$ tetravacancies dominate. Some examples are given in Fig. S1 [21]. At **60 kV** and 500 °C, in TEM mode, we observed the formation of both types of tetravacancies, although it is the

$V_{B1N3}$ defects that predominantly formed, while in STEM mode we have exclusively observed the formation of $V_{B1N3}$ defects. It is worth underlining that in STEM mode the polarity of the defects can be determined directly, while in TEM mode it was assigned by considering that the larger triangular vacancies and edges are N-terminated. As the edge terminations can be influenced by local factors, and in view of previous results showing that B-terminated edges are also favorable [9], this assumption should be made carefully.

Identifying the formation mechanism of the defects is complicated by the fact that there are multiple damage mechanisms which act simultaneously. The minimum acceleration voltage required to create defects in pristine h-BN via **elastic knock-on displacements** of B (N) atoms is 80 kV (120 kV). This value decreases to 55 kV (80 kV) for two coordinated atoms located at the edges of small vacancies [25]. Damage is created in h-BN even when imaging below these thresholds [26] which means that other damage mechanisms, such as **inelastic interactions** or **electron-beam mediated etching** by contaminants, should also be considered.

The formation of the sister $V_{B3N1}$ tetravacancy through elastic effects is easy to understand at 80 kV, where our STEM observations agree with the previous TEM studies. The same logic does not hold for the $V_{B1N3}$ defect, which cannot be explained through knock-on processes. One of the main differences between our high-temperature experiments and the previous work is that we observe the formation of N single-vacancies. Considering the prohibitively-high knock-on threshold, this points to another mechanism which preferentially consumes N. One possible explanation is the chemical reaction with the vacuum residuals or impurities on the sample. As these processes are enhanced at high temperature, it would explain the dependency that we observe.

The fact that vacancy creation and evolution differs as a function of temperature is surprising. We expect similar effects when extending this process to other materials, which could potentially reveal new defect structures.

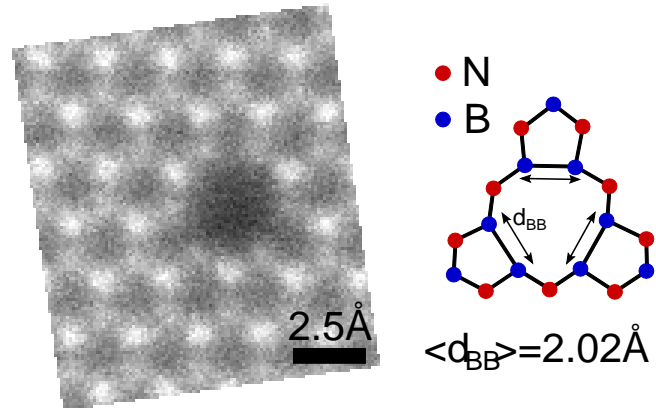

FIG. 1 (color online). High resolution ADF image of a $V_{B1N3}$ tetravacancy in h-BN (left) and corresponding atomic model of the defect (right). Data acquired using 60 kV electrons @ 500 °C.

In order to gain insight into the local properties of the defects, we performed EELS mapping, which showed anomalies in the shape of the B K-edge. This is illustrated in Fig. 2, which shows spectra extracted from a map that was acquired in an area surrounding a tetravacancy. The full map is shown in Fig. S2 [21]. The green spectrum, extracted from the center of the defect, shows distinct features which are not found in other spectra taken farther away from the vacancy, as exemplified by the red spectrum acquired from a similarly-sized area. These features include a shift in the $\pi^*$ peak, which is located around 191.1 eV, and the appearance of second peak around 196.2 eV, between the $\pi^*$ and $\sigma^*$ peaks.

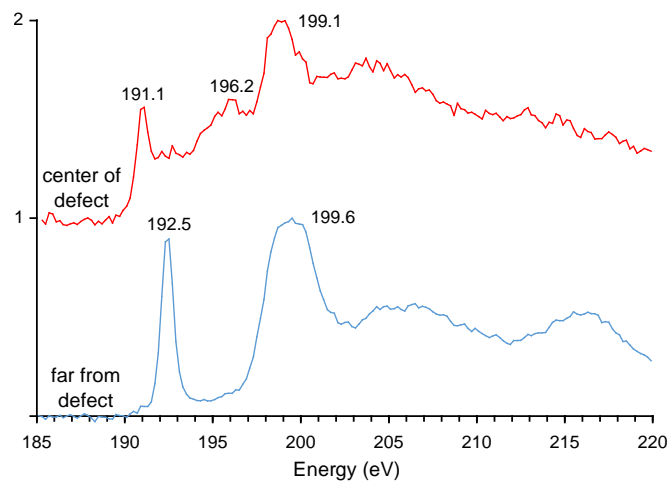

FIG. 2 (color online). The B K-edge EELS signature of a tetravacancy. The two spectra are acquired from the center of the defect and near the defect, respectively. The spectra have been PCA-filtered. Data acquired using 60 kV electrons @ 600 °C.

Further evidence of this spectral anomaly can be found by looking at an EELS map taken from a different ($V_{B1N3}$) tetravacancy, displayed in Fig. 3. Fig. 3(a) shows the ADF image acquired during mapping. An atomic model of the defect is overlaid in Fig. 3(b). PCA filtered maps of the 191.1, 192.5 and 196 eV peaks are displayed in Fig. 3(c)-(e), respectively. There is a striking difference between the 192.5 eV $\pi^*$ peak of B, which is uniformly distributed on the surface of the flake outside the defect and the newly-observed 191.1 and 196 eV peaks, which are localized around the area of the vacancy, confirming that these features are a signature of the atoms that form the defect.

The good agreement between the maps in Fig. 3 and Fig. S2 [21] makes us confident that the EELS signatures that we observe are due to the defects themselves. However, there is always the possibility of contaminants that are trapped around the defect and change the local chemical environment of the structure. While heavier elements can be distinguished by their stronger contrast in the image (Si is typical for our samples), this is challenging to do with those that are close to the sample atoms. In particular, we have checked for the signature of C in several of our EELS maps and found no traces in the mapped area.

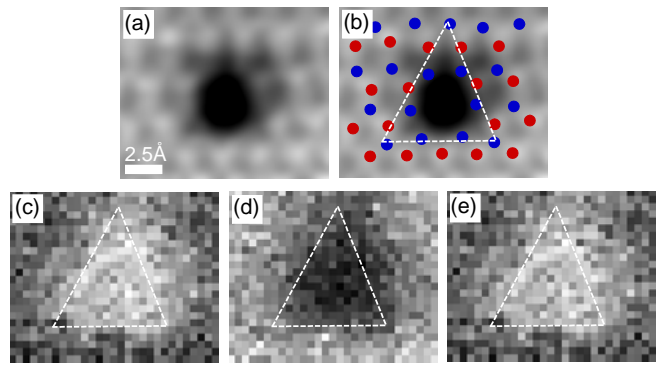

FIG. 3 (color online). EELS mapping of the B K-edge around a tetravacancy. The ADF signal acquired during the scan is shown in (a); the image has been low-pass filtered. An atomic model of the defect is overlaid in (b); the positions of the B and N atoms are marked in blue and red, respectively. PCA-filtered maps of the 191.1, 192.5 and 196 eV peaks are displayed in (c)-(e), respectively. Data acquired using 30 kV electrons @ 500 °C.

In both Figs. 2 and 3, the spectra showing the described anomalies are acquired by placing the beam close to the center of the defect. As one moves closer to the edges of the defect, the EELS signal includes components from several B atoms, due to delocalization. The problem is summarized in [27]; although there is some uncertainty between the various results, the delocalization for the B K-edge (defined as the region containing 50% of the scattered electrons) can be estimated to around 0.3 nm. Under our sampling conditions of about 0.5 Å/px, the delocalization due to sub-scanning is much smaller and can safely be neglected. We analyze this in Fig. S3 [21], by plotting the radially averaged intensities of the peaks in Fig. 3 as a function of distance from the defect center. As expected from the maps, the 191.1 eV peak (corresponding to the defect) decreases with distance, while the 192.5 eV peak (corresponding to the pristine lattice) increases. A third plot integrates the region just beyond the two peaks, which is taken in this case as background. We determine by extrapolation that, in the center of the defect, the first peak should be above the background, while the second one should be just below, resulting in a spectrum that only displays the defect signature, in agreement with the experimental data. This opens the possibility of obtaining the same signature by taking EELS maps of larger defects and open edges, although in our experience these damage too fast in order to allow for this type of analysis at an atomic scale.

In order to understand the origins of the B EELS signature, we have calculated the B K-edge for various atomic positions within and near the tetravacancy. The results are shown in Fig. 4, which compares the experimental data with results from CASTEP calculations. The two calculated spectra plotted in Fig. 4 come from the marked atoms in the displayed model. The data was aligned by shifting the position of the $\pi^*$ peak in the "reference" B atom, which is the farthest from the defect, in order to correspond with the one from the experimental spectrum taken near the defect. The other spectrum was shifted by the same amount, so that there is no artificially-induced offset between the two.

The figure shows very good agreement between the experiment and calculation. The relaxed structure reveals that the distance between the B atoms at the edge of the defect decreases, in agreement with our experimental images and previous calculations [23]. The new 1.99 Å distance is identical, within experimental uncertainties, with the 2.02 Å which we measure. The two calculated EELS signatures also agree well with the experiment. In particular, the data acquired from the center of the defect fits with calculations for the "edge" B atoms located at the edge of the vacancy and proves that they are responsible for the anomalous EELS signature obtained from these regions.

We have additionally calculated the EELS signatures of triple coordinated B atoms near the edge of the defect. The spectra help explain secondary features which are sometimes observed. A summary is given in Fig. S4 [21], which shows a 192.1 eV peak that appears in the data acquired close to the center of the defect which corresponds to the position of the $\pi^*$ peak of the atoms located in the corners of the triangle. A second

192.8 eV peak can be attributed to the nearest-neighbors of the edge B atoms, which show a π* peak which is shifted to 192.65 eV. Separating the various signatures in this area is made difficult by the proximity of the 192.5 eV bulk π* peak, resulting in variations between different positions near the center of the defect, as well as between different PCA filtering algorithms.

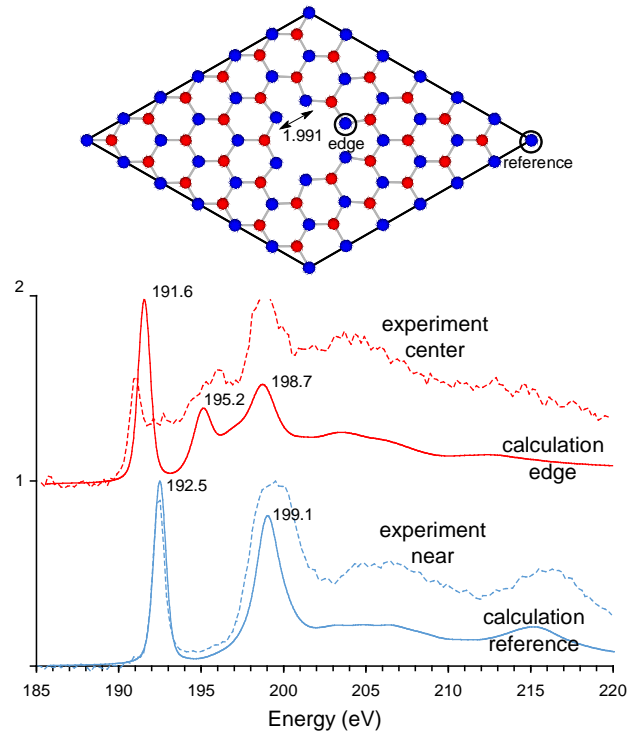

FIG. 4 (color online). Comparison between experimental data and DFT calculations of the B K-edge for atomic positions at the edge and near the tetravacancy. The model used, highlighting the respecting positions, is displayed above.

To summarize, we have used high-resolution imaging and spectroscopy in order to get information about the structure and distribution of chemical properties of boron-terminated tetravacancies in h-BN. We have discussed the formation of this defect and identified key differences between this study and previous reports. We have confirmed that the structure relaxes by reducing the distance between the under-coordinated B atoms, confirming theoretical predictions. We have identified new spectral features for B atoms at the edges of the defect, using high resolution EELS mapping. We have performed calculations which reproduce and help explain the observed features. Our local characterization of these defects is important for understanding the way in which they change the overall properties of this material.

The work is partially supported by the JST Research Acceleration Program. MK acknowledges financial support by JSPS KAKENHI (grant number 23681026 and 26390004). ZL acknowledges support by MEXT KAKENHI Grant Number 25107003.

# Supplementary Information

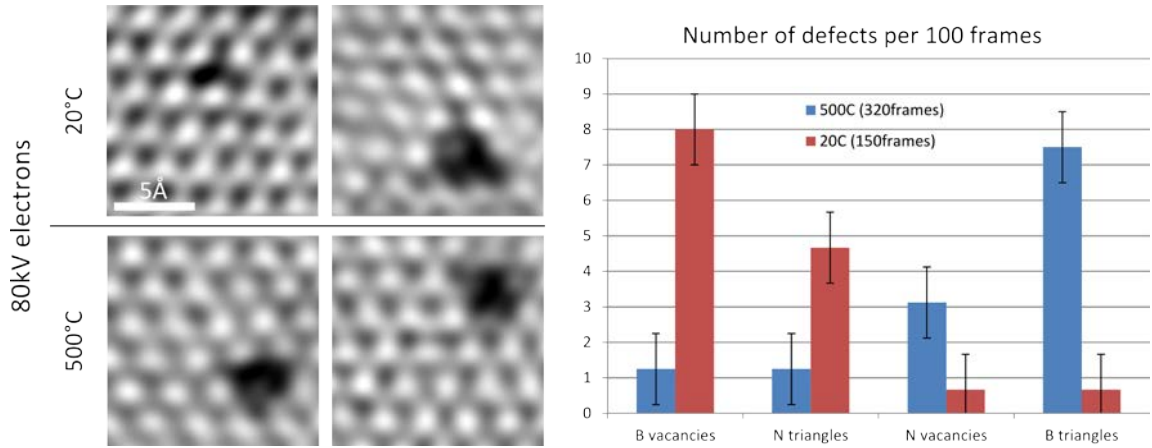

FIG. S1. (Left) ADF images of defects created in h-BN under 80 kV electron irradiation. The images have been FFT-filtered. At room temperature (upper row), B vacancies and $V_{B3N1}$ defects dominate, while at high temperatures (lower low), N vacancies and $V_{B1N3}$ defects are predominant. (Right) Distribution of the various types of defects created at 80 kV as a function of temperature. The error bars have been introduced in order to account for inaccuracies in structure assignment.

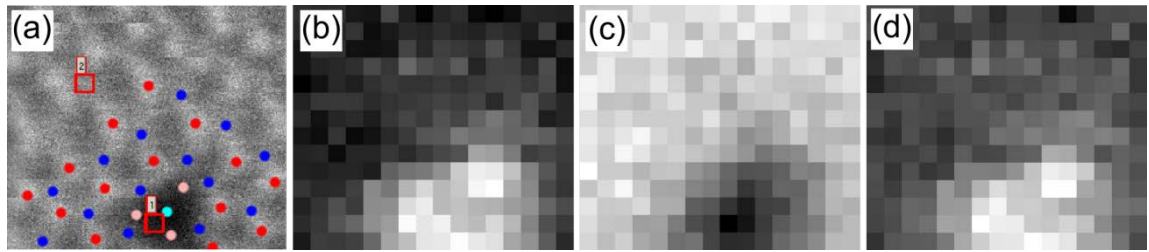

FIG. S2. EELS mapping of the B K-edge around a tetravacancy. The ADF signal acquired during the scan is shown in (a); an atomic model of the defect is overlaid (the positions of the B and N atoms are marked in blue and red; the missing atoms are marked using lighter colors). PCA-filtered maps of the 191.1, 192.5 and 196.0 eV peaks in Fig. 2 are displayed in (b)-(d) respectively. Data acquired using 60 kV electrons @ 600 °C.

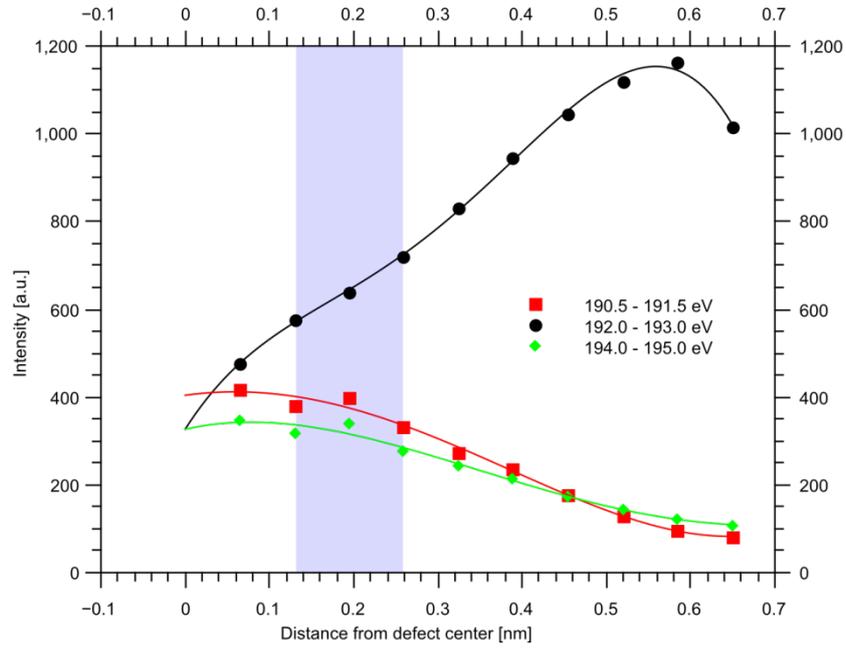

FIG. S3. Radially averaged distribution of intensities for three energy ranges from the EELS map in Fig. 3, as a function of distance from the defect center. The lines represent polynomial fits to the curves. The shaded rectangle highlights the area where the edge atoms are located; the three curves each display local minima or maxima at this position.

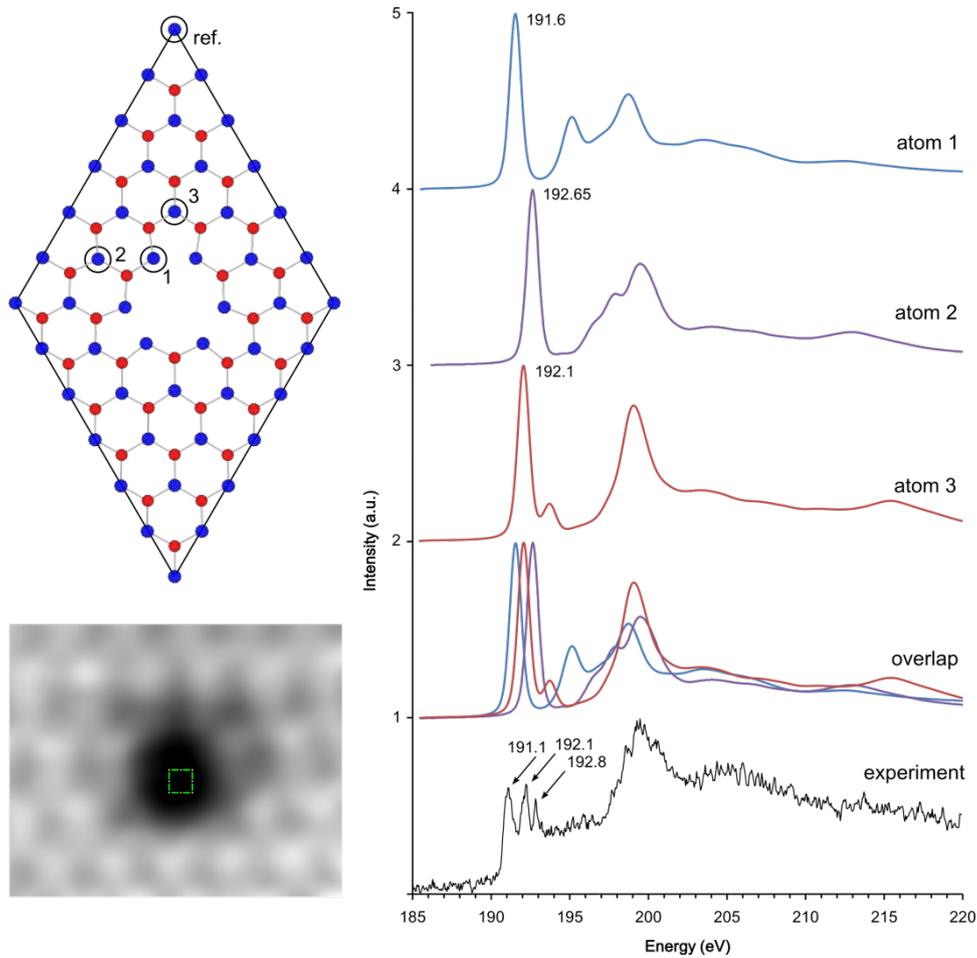

FIG. S4. CASTEP calculations of the B K-edge for three representative atomic positions near the edge of the tetravacancy. The model used, highlighting the respecting positions, is displayed on the left. The calculated data is displayed in the right-hand side panel. An overlap is included in order to make comparison easier. An experimental spectrum showing two additional peaks around 192.1 and 192.8 eV is shown in the same figure. The exact area where the data was acquired is displayed in the lower-left panel. Data acquired using 30 kV electrons @ 500 °C, from the same dataset as Fig.3 in the main text.

**Band structure calculations**

A single layer h-BN sheet was modeled by removing two intermediate BN layers from the crystal structure available at the inorganic materials' structural database [1] and the literature [2], resulting in two layers with a 1 nm separation along the c-axis.

A 6x6x1 supercell was created in order to accommodate the tetravacancy, in which a boron atom with a core-hole is far enough away from its equivalents in order not to interfere with each other, considering the periodic nature of the model. We performed the geometry optimization by using the generalized gradient approximation

(GGA) – PBE functionals [3] with ultra-fine quality. The calculation parameters for ultra-fine quality were: the convergence threshold for the maximum energy change, maximum force, maximum stress and maximum displacement set to $5.0\times10^{-6}$ eV/atom, 0.01 eV/Å, 0.02 GPa, and $5.0\times10^{-4}$ Å, respectively. An SCF tolerance smaller than $5.0\times10^{-7}$ eV/atom was regarded as convergence. The BFGS line search was adopted. We also set 440.0 eV of Energy cutoff, 48×48×48 FFT grid density, and a 1×1×1 K-point set. Either ultrasoft or norm conserving Pseudopotentials represented in a reciprocal space were selected as an "on the fly" option. Density mixing of 0.5 charge was used for electronic minimizer. 20% of empty bands with 0.1 eV of smearing were considered for orbital occupancy.

For Energy estimation of core level spectroscopy, we adopted the "ultra-fine" presets for SCF iterations with the following parameters: generalized gradient approximation (GGA) – PBE functionals [3] (1×1×1) K-point set, $5.0\times10^{-7}$ eV/atom SCF tolerance, 440 eV plane wave basis set cut-off energy and 48×48×48 FFT grid density with an augmentation density scaling factor of 1.5. Proper electronic minimization parameters (20 % of empty bands and 0.1 eV of smearing), instead of assigning fixed occupancy, were required for SCF convergence.

The shapes of boron K-edge ELNES from each atomic site are simulated by the "core spectroscopy" function in CASTEP after self-consistent field (SCF) iterations for the final state (core-hole) electron configurations. The SCF iterations were converged based on GGA – PBE functionals [3] with ultra-fine quality, similar to the geometry optimization. The obtained spectra are smeared with 0.3 eV of energy broadening (FWHM) with a consideration of lifetime effect, $0.3 + 0.3(\varepsilon_k - \varepsilon_{threshold})$ eV, where the FWHM increases with increasing energy difference between the threshold energy, $\varepsilon_{threshold}$, and the energy above the threshold, $\varepsilon_k$ [4]. A core-hole was introduced by removing an electron from core level at the atom of interest.

The estimation of threshold energy based on pseudo potential method by Mizoguchi et al. contains 1 % of uncertainty [5]. In the pseudo potential method, the difference in total energy from the final state electron configurations (core-hole) to the ground state electron configuration (without core-hole) is regarded as the energy difference in the valence state, $\Delta E_{valence} = \Delta E_{final} - \Delta E_{ground}$. Here, we assume the energy of bulk boron having a π* peak at 192.5 eV as a threshold energy, $E_{TE}$, and we used the reference to estimate the threshold energy of different edge structures. Based on the assumption, we can obtain the energy difference of the unknown core state of bulk boron atoms, from the equation $E_{TE}$ (bulk boron=192.5 eV) = $\Delta E_{valence}$ (bulk boron) + $\Delta E_{core}$ (bulk boron). The threshold energy of the edge boron or other types of boron can be also estimated from the equation, $E_{TE}$ (edge boron) = $\Delta E_{valence}$ (edge boron) + $\Delta E_{core}$ (edge boron). Here we also assume that the energy difference in core state is identical between bulk boron and edge boron, $\Delta E_{core}$ (bulk boron) = $\Delta E_{core}$ (edge boron).